# Ordered creation and motion of skyrmions with surface acoustic wave

Ruyi Chen[#], Chong Chen[#], Lei Han[#], Peisen Liu, Rongxuan Su, Wenxuan Zhu, Yongjian Zhou, Feng Pan, Cheng Song*

Key Laboratory of Advanced Materials (MOE), School of Materials Science and Engineering, Beijing Innovation Center for Future Chip, Tsinghua University, Beijing 100084, China

**Magnetic skyrmions with a well-defined spin texture have shown unprecedented potential for various spintronic applications owning to their topologically non-trivial and quasiparticle properties. To put skyrmions into practical technology, efficient manipulation, especially the inhibition of skyrmion Hall effect (SkHE) has been intensively pursued. In spite of the recent progress made on reducing SkHE in several substituted systems, such as ferrimagnets and synthetic antiferromagnets, the organized creation and current driven motion of skyrmions with negligible SkHE in ferromagnets remain challenging. Here, by embeding the [Co/Pd] multilayer into a surface acoustic wave (SAW) delay line, we experimentally realized the ordered generation of magnetic skyrmions. The resultant current-induced skyrmions movement with negligible SkHE was observed, which can be attributed to the energy redistribution of the system during the excitation of SAW. Our findings open up an**

[#]These authors contributed equally: Ruyi Chen, Chong Chen, Lei Han
*E-mail: songcheng@mail.tsinghua.edu.cn

**unprecedentedly new perspective for manipulating topological solitons, and would advance the development of skyrmionics and spin acousto-electronics.**

Efficient generation and manipulation of spin textures such as magnetic skyrmions have been a long-standing theme in the field of spintronics for their various potential applications, such as information storage[1–3], logic computing gates[4,5] and neuromorphic computing devices[6,7]. In ferromagnetic system, magnetic skyrmion was initially reported in chiral itinerant-electron magnet MnSi at low temperature[8]. Later, various methods, such as magnetic field[9–11], electric current/field[12–14], and thermal gradient[15–19] were proposed to generate skyrmions in magnetic multilayers with interfacial Dzyaloshinskii-Moriya interaction (DMI) at room temperature. Nevertheless, skyrmions generated in such studies are randomly distributed, making it problematic to be utilized as information carriers. On the other hand, SkHE is another obstacle that restricts the essential transmission of skyrmions in devices, where skyrmions feel the Magnus force due to the finite topological charge[3,20,21]. Such Magnus force deflects the movement trajectory of magnetic skyrmions from the driving current, which is unfavorable for the storage stability and further device reliability. Although several material systems, such as ferrimagnets[22–24] and synthetic antiferromagnets[25–27] have been demonstrated to possess skyrmions with reduced SkHE, the organized creation of skyrmions and current-driven motion with negligible SkHE in the prototypical ferromagnets still remain challenging, which has tremendously limited the applicability of skyrmions in practical devices.

A promising strategy for modifying magnetic textures at the nanoscale together with high controllability is to use strain. Surface acoustic waves are typical strain waves that can be excited through oscillating electric fields and propagate millimeter distances at the surface of piezoelectric materials[28,29]. By depositing magnetic films on the surface, the alternating strain generated in them can modify the magnetic interactions through the magnetoelastic effect, making it possible for manipulating magnetic states[30]. Indeed, SAWs have already been used to induce the magnetization oscillations[31,32], to assist the switching of the magnetic moments[33,34], and to control the dynamics of magnetic textures[30,35,36]. Here, by fabricating an integrated SAW device, we report an experimental realization of the organization of magnetic skyrmions and the resultant current-induced skyrmions movement with negligible SkHE, simultaneously. We note that the comb-shaped interdigital transducers (IDTs) used in the measurements not only can generate thermal effect to assist the formation of magnetic skyrmions but can also excite the SAWs to induce alternating strain in magnetic films. In particular, the energy redistribution of the system caused by the strain gradient provides an effective tool to manipulate skyrmions. Hence, under the excitation of SAWs, skyrmions are initially created with a random distribution and then pushed towards the anti–nodes of the waves, exhibiting an ordered feature. Furthermore, theoretical simulations were performed to explain this pinning of skyrmions at the anti-nodes of SAWs through the view of energy variation. Our results complement the efficient manipulation of magnetic skyrmions in ferromagnets and may advance the further investigation of technologically relevant physics/device conceptions.

We start by discussing our approach to realize the organized motion of magnetic skyrmions in ferromagnets embedded into the SAW delay line. Two aspects of issues need to be prepared in our scenario. On one hand, the sample design, such as the magnetic multilayers with moderate perpendicular magnetic anisotropy (PMA), interfacial DMI, and film thickness play a significant role on the generation of magnetic skyrmions[37]. Here, the Co/Pd/Co/Pd/Co/Pt multilayer structure was chosen in our experiments not only for the proper magnetic parameters and small dipolar fields in favour of skyrmion stability[11] but also for their potential to generate skyrmions by thermal effect[19]. On the other hand, to make use of the SAWs-induced magnetoelastic effect, we fabricated the delay line device which consists of a two-port IDTs with a magnetic channel embedded in the cavity as shown in Fig. 1a. Here, a wire geometry was designed for the following dynamic measurements. To create magnetic skyrmions, a radio-frequency (RF) voltage was inputted into IDTs to excite the SAWs together with thermal effect. It is noted that since the strain gradient induced by SAWs is periodic with corresponding force vanishing at the anti-nodes of the wave[30], the generated skyrmions by thermal effect exhibit a more stable state at the anti-nodes of SAWs. Thus, by means of magneto-optical Kerr effect (MOKE) microscopy, we can directly observe the pinning of skyrmions in magnetic film with an ordered alignment. In addition, the transversal component during the current-driven motion of skyrmions should also be suppressed by SAWs, making it promising to eliminate the SkHE. Stack structure of Co(0.3)/Pd(0.9)/Co(0.3)/Pd(0.9)/Co(0.3)/Pt(1.4) (units in nanometer) was deposited on 128°-rotated, Y-cut LiNbO$_3$ substrate via magnetron sputtering (Methods). In Fig. 1b, we present the temperature distribution image of the integrated device obtained by infrared camera. The width and gap of the fingers are both designed as 5 μm to excite a SAW with a well-defined wavelength, and the width of the magnetic channel is designed as 60 μm for the current-induced skyrmions movement measurements. By applying RF voltage with 365.65 MHz and 21 dBm to the IDT, the excitation of SAWs lead to the apparent increase of temperature which is favorable for the generation of skyrmions. Figure 1c shows the typical SAW reflection spectrum of the delay line, which contains three excited modes corresponding to the first (181.63 MHz), second (365.65 MHz) and third (546.03 MHz) harmonic. The transmission spectra between the two IDTs are also presented in the insert of Fig. 1c. Three peaks appeared in transmission spectrum match well with the reflection spectrum, determined by the sound velocity of the piezoelectric substrate and the geometry of IDTs. The similar features of the transmission spectrum are also observed in the finite element simulations. (Supplementary Fig. S3) We next estimate the thermal effect during the excitation of SAWs using the infrared camera. In Fig. 1d, we summarized the average temperature at the center of magnetic channel with respect to the three typical frequencies as a function of applied power. It is clear that the temperature in the magnetic films increases from 28.6 °C to about 38.6 °C as the increase of applied power from 17 dBm to 22 dBm at 365.65 MHz. In contrast, the temperatures in the magnetic film change much smaller (below 1 °C) at the same range of power at 181.63 MHz and 546.03 MHz. Therefore, we note that the large dip at the second harmonic provide both the SAWs and thermal effect in our delay line, paving the way for the later creation of organized skyrmions.

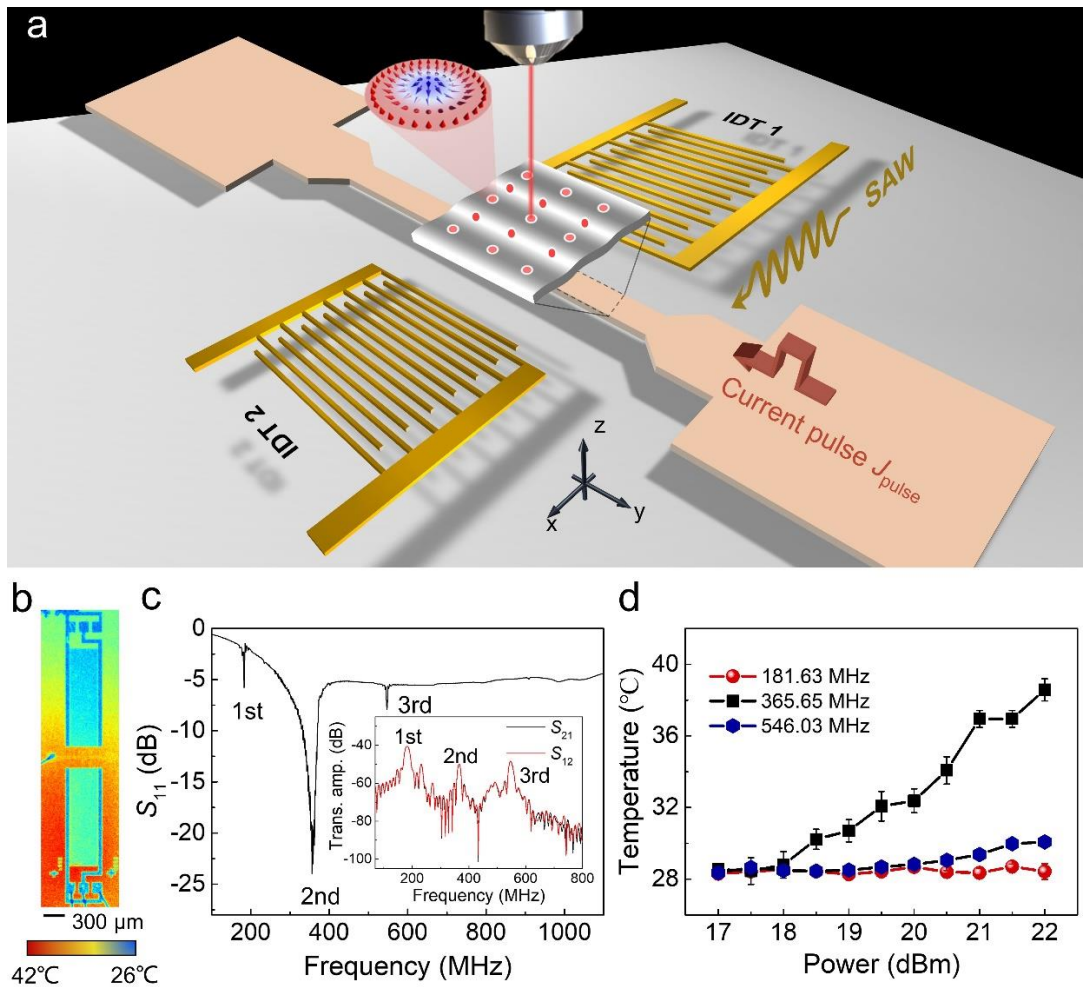

**Fig. 1| Schematics of experimental setup and basic properties of surface acoustic wave devices. a**, Schematic of the integrated device with a pair of IDTs on both sides of the magnetic multilayer channel. Néel type skyrmions are generated at the antinode of SAW

which can be observed via magneto-optical Kerr effect microscopy. Current pulses are applied to drive the movement of created skyrmions along the channel. **b**, Temperature image of the integrated device obtained by the infrared camera when RF voltage of 365.65 MHz and 21 dBm are applied to the IDT. **c**, Typical SAW reflection spectrum of the 5 μm wide delay line device which shows three excited modes corresponding to the first (181.63 MHz), second (365.65 MHz) and third (546.03 MHz) harmonic. Inset is the corresponding transmission spectra with the magnetic channel placed between the two IDTs. All the spectra are obtained using RF signal of –5 dBm. **d**, The average temperature at the center of magnetic channel under different frequencies as a function of applied power. The error bars correspond to the standard deviation.

To validate our design, we next studied the evolution process of magnetic domains in the deposited multilayers with MOKE microscopy. We show in Fig. 2a the consecutive MOKE images acquired from the Co/Pd/Co/Pd/Co/Pt multilayer under different perpendicular magnetic fields with (upper row) and without (bottom row) SAWs. For the applied RF voltage of 365.65 MHz and 21 dBm along $x$-direction, intertwined maze domains were observed at the magnetic field of –2.33 Oe which evolve into the mixture of stripe domains and skyrmions with increasing perpendicular magnetic field. The most striking feature is that all of the generated skyrmions and maze domains align in the $y$-direction, exhibiting an ordered character. On the contrary, for the frequency of 181.63 MHz and 546.03 MHz, no skyrmion or maze domain appears during varying the magnetic field. (Supplementary Fig. S4) As comparison, in the bottom row, only the coherent domain reversal was observed as we scanned the magnetic field from –14.61 Oe, to 25.76 Oe without SAW. Thus, all of the results above indicate that (i) the generation of magnetic skyrmions is the synergetic role of thermal effect and SAWs, and (ii) the ordered alignments of skyrmions are induced by the SAWs. Moreover, the phase diagram of magnetic films is further studied with and without SAWs. The density of skyrmions ($n_{sk}$) as a function of the applied power $P$ under the magnetic field of –12.96 Oe is shown in Fig. 2b. The skyrmions density increases gradually from 0 to $10 \times 10^5$ mm$^{-2}$ with increasing $P$ from 19 to 23 dBm. A skyrmions density phase diagram summarizing the evolution of coexisting stripe domain and skyrmions, maze domain, and uniformed domain states as a function of the magnetic field with and without SAWs is shown in Fig. 2c. Three different magnetic phases can be distinguished under the excitation of SAWs scenario. When the magnetic fields are small ($|H_z| < 4$ Oe), only maze domain can be observed, which is due to the local instability of skyrmion phase under weak magnetic fields. When 4 Oe $< |H_z| < 21$ Oe, the maze domain evolves into the mixture of skyrmions and stipe domains and the skyrmions density experience the first rises and then falls as increasing the magnetic fields. This can be attributed to the reason that the increasing Zeeman energy first overcomes the energy barrier to nucleate skyrmions and stripe domains and then destroies their stability as it is large enough. Finally, when the magnetic fields are large (21 Oe $< |H_z|$), the whole system transform into the uniform states and skyrmions disappear resulted from the forbidden of skyrmions by large Zeeman energy in the energetically stable ferromagnetic states[38]. In comparison, no skyrmion is observed in the film during scanning the magnetic field without SAW, indicating the necessary of SAW for the creation of magnetic skyrmions.

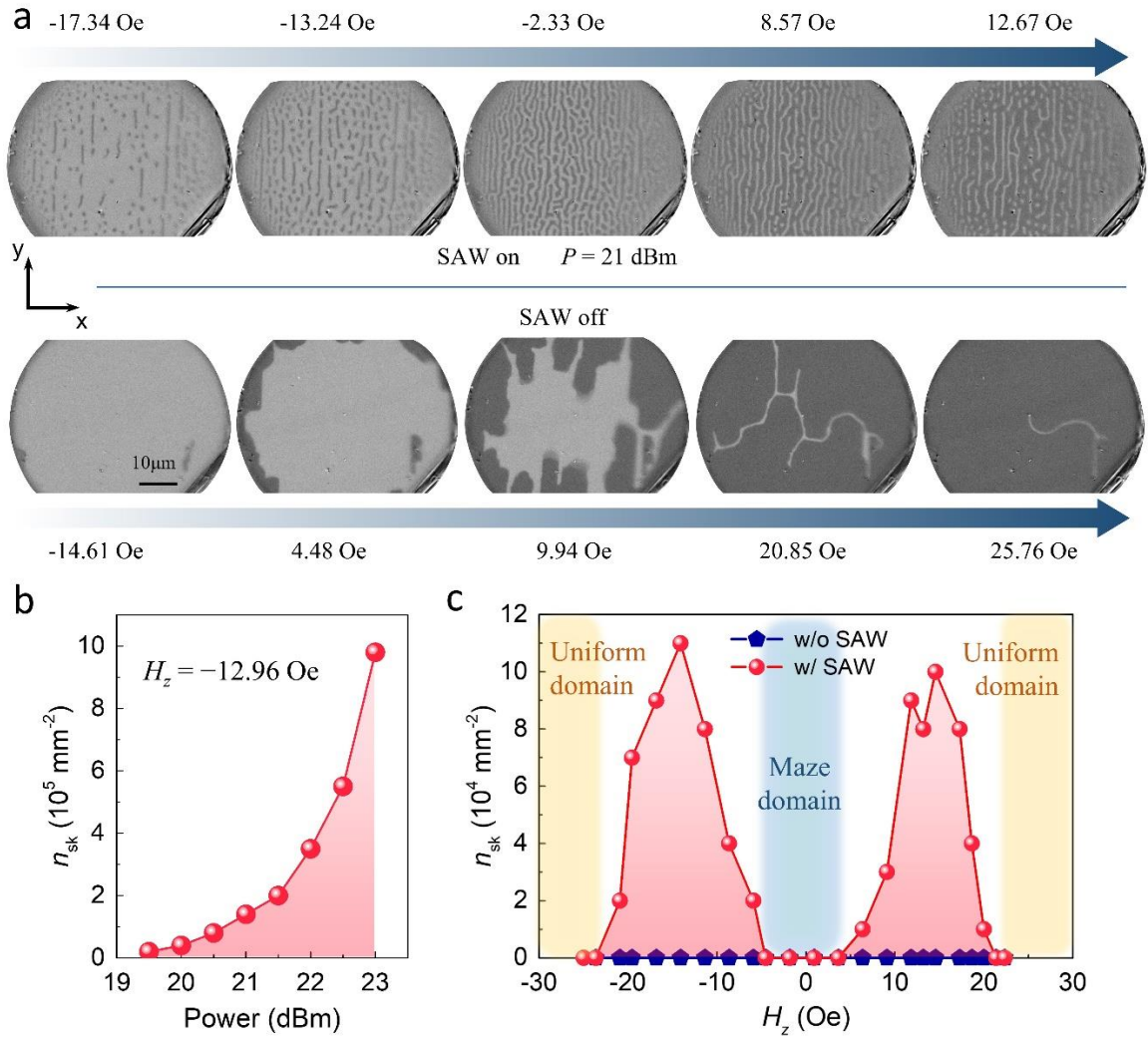

**Fig. 2| Transformational dynamics and phase diagram of magnetic films with and without the excitation of SAWs. a,** Consecutive MOKE images acquired from the Co/Pd/Co/Pd/Co/Pt multilayer under different perpendicular magnetic fields with (upper

row) and without (bottom row) SAWs. Images taken at –17.34 Oe, –13.24 Oe, –2.33 Oe, 8.57 Oe and 12.67 Oe with the RF signal of $P = 21$ dBm are shown in the upper column. Images taken at –14.61 Oe, 4.48 Oe, 9.94 Oe, 20.85 Oe and 25.76 Oe without SAW are shown in the bottom column. **b**, The density of skyrmions ($n_{sk}$) as a function of the applied power $P$ under the magnetic field of –12.96 Oe. **c**, The density of skyrmions phase diagram summarizing the evolution of different magnetic phase as a function of magnetic field $H_z$ with and without SAWs.

We now turn to the dynamic behavior of these orderly created skyrmions to see whether SAWs can suppress the transversal movement of current-induced skyrmions motion. Figure 3a displays a series of sequential MOKE images of magnetic domain patterns of Co/Pd/Co/Pd/Co/Pt multilayer under the current pulse of $2.5 \times 10^{10}$ A m$^{-2}$ at $H_z = 10.86$ Oe. The RF signal of $P = 21$ dBm was applied during the current-driven experiments. Here, different skyrmions are featured by different colors for the sake of distinction. Apparently, skyrmions move along the same direction of current pulses, confirming that these skyrmions are topologically protected with chiral Néel-type configuration[11] and their motion are governed by spin-orbit-torque originated from the heavy layer Pt[26]. Strikingly, these skyrmions move almost in a straight line with negligible transverse component according to the movement trajectory presented in Fig. 3b, suggesting the reduced SkHE in our experiment. Moreover, to quantitative demonstrate the influence of SAWs on SkHE, the current-driven skyrmion Hall angle $\theta_{sk} = \tan^{-1}(v_x/v_y)$ is evaluated, where $v_x$ and $v_y$ are velocity in the $x$ and $y$ directions[21]. In Fig. 3c, we summarize the phase diagram of the skyrmion Hall angle $\theta_{sk}$ as a function of current density with and without SAWs for skyrmions with topological charge $Q = +1$ and $Q = -1$. The skyrmions in the sample without SAW are generated through the thermal effect by fabricating a heater on the side of the channel (Supplementary Fig. S5). When the current density is small ($J < 1.25 \times 10^{10}$ A m$^{-2}$), skyrmions cannot be driven by the current for neither SAW nor the thermal induced cases. This is due to the fact that a certain large of current is needed to overcome the skyrmions pinning barrier. As increasing current density from $1.25 \times 10^{10}$ A m$^{-2}$ to $10 \times 10^{10}$ A m$^{-2}$, skyrmions start to move and the corresponding skyrmion Hall angle increases rapidly from 0° to around 30° for the sample with thermal effect alone. In comparison, the $\theta_{sk}$ maintains at a small level (< 7 °) for the sample with SAWs. Such a distinct difference of these two scenarios suggests that SAWs play a significant role on the suppression of transversal movement during current-induced skyrmion motion which shows considerable potential to inhibit the SkHE in ferromagnets.

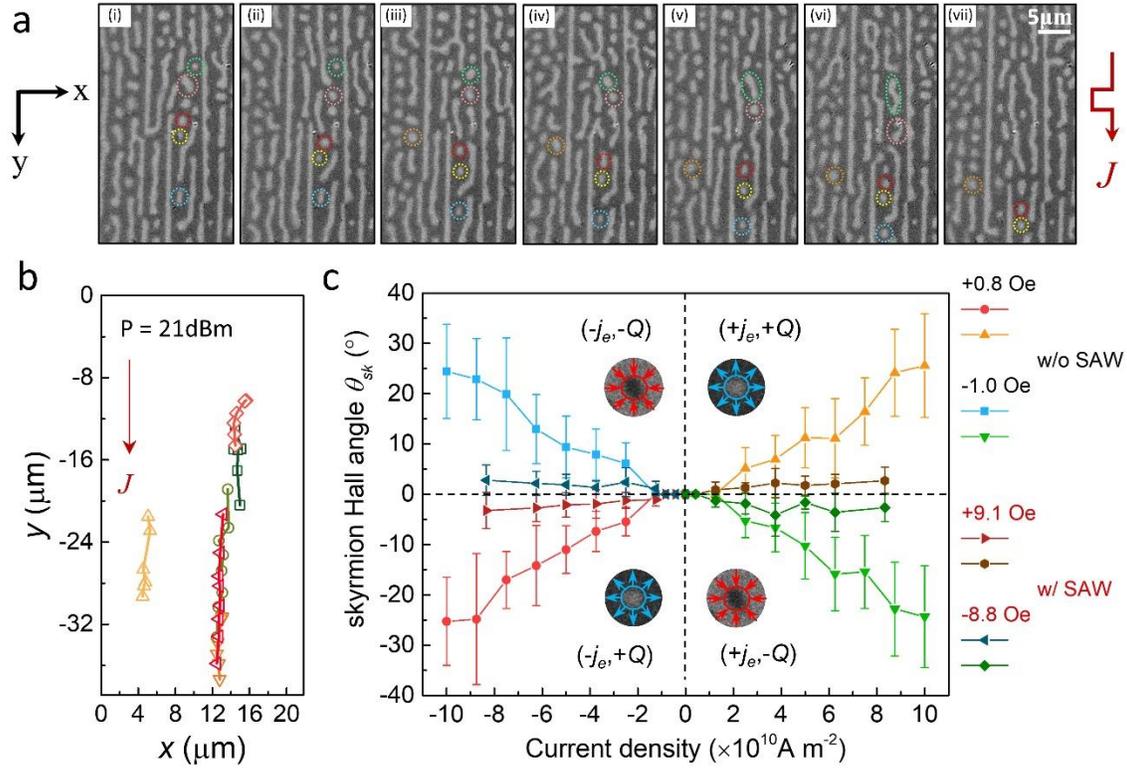

**Fig. 3| Current-induced behavior of magnetic skyrmions and their skyrmion Hall effect with and without the excitation of SAW. a**, Sequential MOKE images of magnetic domain patterns of Co/Pd/Co/Pd/Co/Pt multilayer under the current pulse applications at $H_z$ = 10.86 Oe. Note that different skyrmions are characterized by different colors. Current density used to drive the skyrmion movement is $2.5 \times 10^{10}$ A m$^{-2}$. **b**, The movement trajectory of selected skyrmions in (**a**). **c**, Phase diagram of the skyrmion Hall angle $\theta_{sk}$ as a function of current density measured with and without SAWs for skyrmions with topological charge $Q = +1$ and $Q = -1$. The skyrmions in the sample without SAW are generated through the thermal effect by fabricating a heater on the side of the channel. Each data point is an average of measurements obtained by tracking the movement of several independent skyrmions, with the error bars representing the standard deviation. The insets

show the magnetic configurations of the skyrmions with opposite topological charge. Note that $H_z = +9.1$ Oe ($-8.8$ Oe) for $Q = +1$ ($-1$) with SAWs and $H_z = +0.8$ Oe ($-1.0$ Oe) for $Q = +1$ ($-1$) without SAW. The RF signal of $P = 21$ dBm was applied during all the current-driven measurements.

To discuss the microscopic origin of these experimental observations, we further carried out micromagnetic simulations in a quasi-two-dimensional system and studied the evolution process of spin textures. The magnetoelastic coupling energy can be described as[39–42]:

$$E = b_1 \sum_i m_i^2 \varepsilon_{ii} + b_2 \sum_{i \neq j} m_i m_j \varepsilon_{ij}$$

where $b_1$ and $b_2$ are the magnetoelastic coefficients, and $\varepsilon_{ij}$ is the strain tensor induced by SAW. In a Rayleigh mode, the only existing strain components are $\varepsilon_{xx}$, $\varepsilon_{xz}$ and $\varepsilon_{zz}$, where $\varepsilon_{xz}$ is phase shifted by 90° with respect to $\varepsilon_{xx}$[43]. Hence, the magnetoelastic coupling energy can be simplified as[41,44]:

$$E = b_1 \varepsilon_{xx} m_x^2 + b_1 \varepsilon_{zz} m_z^2 + 2b_2 \varepsilon_{xz} m_x m_z$$

In Fig. 4a, we present the magnetization $m_z$ distribution of ferromagnetic layer after the excitation of SAW where both skyrmions and stripe domains are aligned in an ordered way along the $y$-direction. Note that partial skyrmions and stripe domains are not perfectly aligned which may be ascribed to the existence of self-interactions between skyrmions and stripes. In fact, this phenomenon is also observed in our experiments as illustrated in Fig.

2a. For a deeper understanding, we then studied the entire evolution process of the selected area in Fig. 4a and the colour maps of the strain, magnetization $m_z$, magnetization $m_x$ and SAW energy density at different simulation time ($t$) from 0.01 ns to 5.89 ns are shown in Fig. 4b-q. First, from 0.01 ns to 3.86 ns (without SAW), the film relaxes from a multi-domain state (Fig. 4f) to the skyrmion state (Fig. 4g) under the role of perpendicular magnetic anisotropy, interfacial DMI, external magnetic field, and the thermal fluctuations. Apparently, the generated skyrmions here are randomly distributed without any specific rules. Then, at $t = 5$ ns, when the SAW was excited, spatially organized SAW energy density distribution appears (Fig. 4p and 4q). This organized SAW energy density is generated by the magnetoelastic coupling effect that varies with the alternating strain $\varepsilon_{xx}$. Therefore, the SAW related energy at skyrmion center can be written as $E_{skyrmion} = b_1 \varepsilon_{zz} m_z^2$, which exhibits a more stable state at the anti-nodes compared with the nodes of the wave. In this way, the generated skyrmions gradually move toward the anti-node of the wave and finally constitute an organized alignment (Fig. 4i). To analyze the change in energy, we present in Fig. 4r and Fig. 4s the total energy and SAW energy vary as a function of simulation time, respectively. Before exciting SAW, the total energy decreases as the domains evolve from the initial state to the stable skyrmions state. Then, with the applied SAW, the total energy tends to reach a new state which varies with the oscillation of SAW energy (Fig. 4s). In other words, the emergent SAW redistributes the total energy of the system and makes skyrmions more stable at the anti-nodes of wave, leading to their ordered alignment.

Another evidence is that we also carried out the similar experiment in a delay line with the SAW frequency of 2.79 GHz while the created skyrmions are randomly distributed without ordered character. This may due to the wavelength of SAW is so small compared with the skyrmion size that there is no anti-node available for skyrmions alignment (Supplementary Fig. S6).

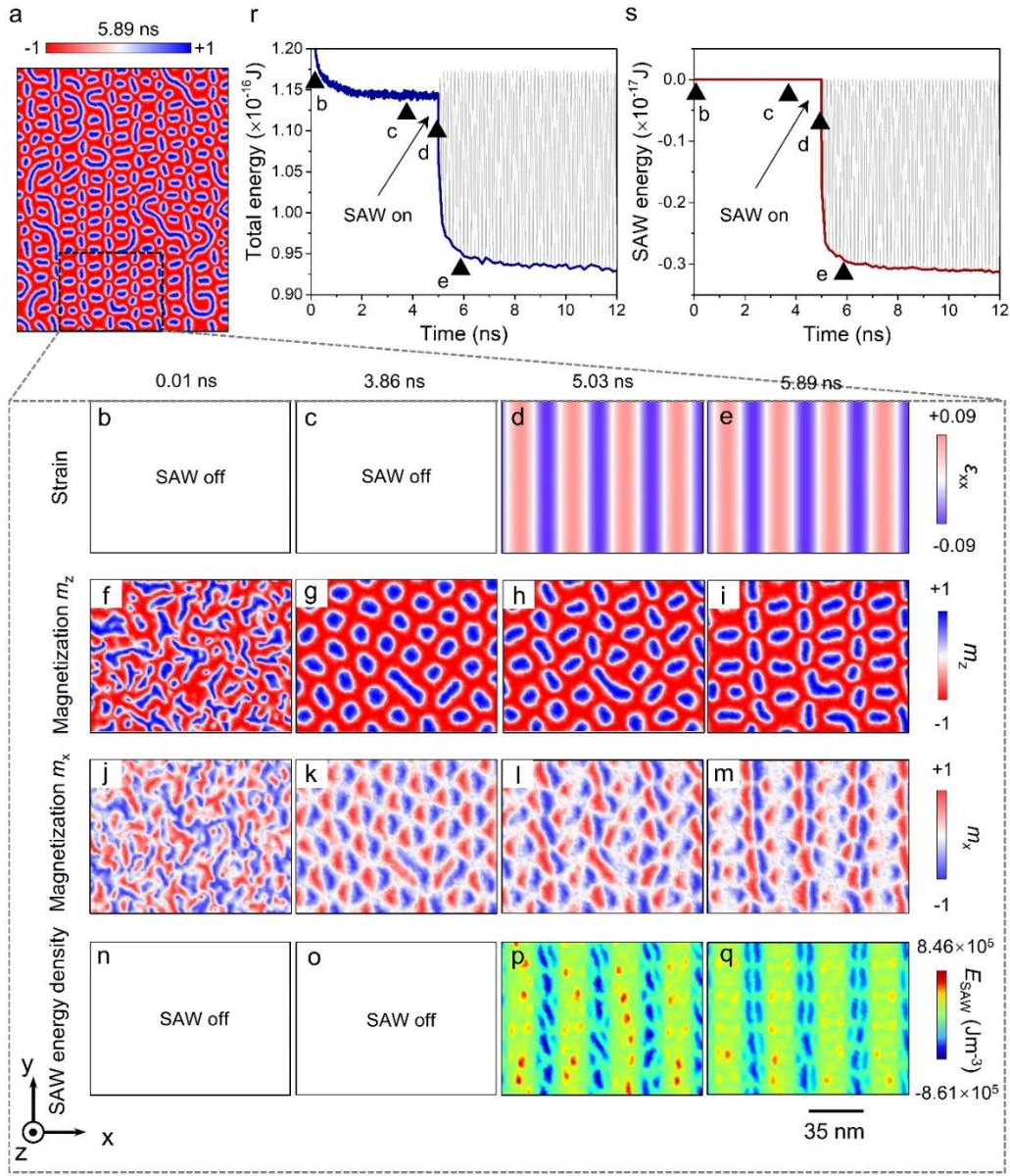

**Fig. 4| Micromagnetic simulations of the SAW induced skyrmions alignment. a,** The magnetization distribution of the simulated ferromagnetic layer at 5.89 ns. Here, the SAW was excited at 5 ns. The generated skyrmions and stripe domains are aligned in an organized way under the excitation of SAW. **b–q,** Colour maps of the strain $\varepsilon_{xx}$ (**b–e**), magnetization $m_z$ (**f–i**), magnetization $m_x$ (**j–m**) and SAW energy density (**n–q**) for the selected area in (**a**) at different simulation time from 0.01 ns to 5.89 ns. **r, s,** The total energy (**r**) and SAW energy (**s**) vary with the simulation time. The solid triangles represent the times corresponding to the colour maps of **b–e.**

In conclusion, we have demonstrated the ordered generation, manipulation and current-driven dynamics of magnetic skyrmions by the integrated SAW devices. We demonstrate that two aspects of issues need to be satisfied to realize the orderly alignment and movement of skyrmions. One is the magnetic structure design that we chose the Co/Pd/Co/Pd/Co/Pt multilayers with proper PMA, interfacial DMI, and film thickness. The other is the integrated delay line structure which can provide SAWs and thermal effect simultaneously. Thus, owning to the energy redistribution derived from the strain gradient during the excitation of SAWs, magnetic skyrmions are confined at the anti-nodes of the SAW and move nearly in a straight line under the current pulses. The unprecedented reduction of SkHE in ferromagnets is expect to drive the progress of skyrmion-based applications. Besides, our findings provide a completely new approach to manipulate topological solitons, which may advance the progress of further skyrmion-based spintronic devices.

**Online content**

Any methods, additional references, Nature Research reporting summaries, source data, supplementary information, acknowledgements, peer review information; details of author contributions and competing interests; and statements of data availability are available at https

**References**

1. Parkin, S. S. P., Hayashi, M. & Thomas, L. 2008 Magnetic domain-wall racetrack memory. *Science* **320**, 190−194 (2008)

2. Fert, A., Cros, V. & Sampaio, J. Skyrmions on the track. *Nat. Nanotechnol.* **8**, 152−156 (2013).

3. Nagaosa, N. & Tokura, Y. Topological properties and dynamics of magnetic skyrmions. *Nat. Nanotechnol.* **8**, 899−911 (2013).

4. Zhang, X., Ezawa, M., & Zhou, Y. Magnetic skyrmion logic gates: conversion, duplication and merging of skyrmions. *Sci. Rep.* **5**, 9400 (2015).

5. Upadhyaya, P., Yu, G., Amiri, P. K. & Wang, K. L. Electric-field guiding of magnetic skyrmions. *Phys. Rev. B* **92**, 1344 11 (2015).

6. Huang, Y., Kang, W., Zhang, X., Zhou, Y. & Zhao, W. Magnetic skyrmion-based

synaptic devices. *Nanotechnology* **28**, 08LT02 (2017).

7. Song, K. M. et al. Magnetic skyrmion artificial synapse for neuromorphic computing. *Nat. Electron.* **3**, 148–155 (2020).

8. Mühlbauer, S. et al. Skyrmion lattice in a chiral magnet. *Science* **323**, 915–919 (2009).

9. Heinze, S. et al. Spontaneous atomic-scale magnetic skyrmion lattice in two dimensions. *Nat. Phys.* **7**, 713–718 (2011).

10. Yu, X. Z. et al. Near room-temperature formation of a skyrmion crystal in thin-films of the helimagnet FeGe. *Nat. Mater.* **10**, 106–109 (2011).

11. Yu, G. et al. Room-temperature creation and spin-orbit torque manipulation of skyrmions in thin films with engineered asymmetry. *Nano Lett.* **16**, 1981–1988 (2016).

12. Schulz, T. et al. Emergent electrodynamics of skyrmions in a chiral magnet. *Nat. Phys.*, **8**, 301–304 (2012).

13. Ma, C. et al. Electric field-induced creation and directional motion of domain walls and skyrmion bubbles. *Nano Lett.* **19**, 353–361 (2019).

14. White, J. S. et al. Electric-field-induced Skyrmion distortion and giant lattice rotation in the magnetoelectric insulator $Cu_2OSeO_3$. *Phys. Rev. Lett.* **113**, 107203 (2014).

15. Lin, S. Z., Batista, C. D., Reichhardt, C. & Saxena, A. ac current generation in chiral magnetic insulators and Skyrmion motion induced by the spin Seebeck effect. *Phys. Rev. Lett.* **112**, 187203 (2014).

16. Kong, L.& Zang, J. Dynamics of an insulating Skyrmion under a temperature gradient.

*Phys. Rev. Lett.* **111**, 067203 (2013).

17. Mochizuki, M. et al. Thermally driven ratchet motion of a skyrmion microcrystal and topological magnon Hall effect. *Nat. Mater.* **13**, 241–246 (2014).

18. Wang, Z. D. et al. Thermal generation, manipulation and thermoelectric detection of skyrmions. *Nat. Electron.* **3**, 672–679 (2020).

19. Chen, R. Y. et al. Controllable generation of antiferromagnetic skyrmions in synthetic antiferromagnets with thermal effect. *Adv. Funct. Mater.* **32**, 2111906 (2022).

20. Litzius, K. et al. Skyrmion Hall effect revealed by direct time-resolved X-ray microscopy. *Nat. Phys.* **13**, 170–175 (2017).

21. Jiang, W. et al. Direct observation of the skyrmion Hall effect. *Nat. Phys.* **13**, 162–169 (2017).

22. Hirata, Y. et al. Correlation between compensation temperatures of magnetization and angular momentum in GdFeCo ferrimagnets. *Phys. Rev. B:Condens. Matter Mater. Phys.* **97**, 220403 (2018).

23. Woo, S. et al. Current-driven dynamics and inhibition of the skyrmion Hall effect of ferrimagnetic skyrmions in GdFeCo films. *Nat. Commun.* **9**, 959 (2018).

24. Caretta, L. et al. Fast current-driven domain walls and small skyrmions in a compensated ferrimagnet. *Nat. Nanotechnol.* **13**, 1154–1160 (2018).

25. Legrand, W. et al. Room-temperature stabilization of antiferromagnetic skyrmions in synthetic antiferromagnets. *Nat. Mater.* **19**, 34–42 (2020).

26. Dohi, T., DuttaGupta, S., Fukami, S. & Ohno, H. Formation and current-induced motion of synthetic antiferromagnetic skyrmion bubbles. *Nat. Commun.* **10**, 5153. (2019).

27. Chen R. Y. et al. Realization of isolated and high-density skyrmions at room temperature in uncompensated synthetic antiferromagnets. *Nano Lett.* **20**, 3299–3305 (2020).

28. Weiler, M. et al. Elastically driven ferromagnetic resonance in nickel thin films. *Phys. Rev. Lett.* **106**, 117601 (2011).

29. Foerster, M. et al. Direct imaging of delayed magneto-dynamic modes induced by surface acoustic waves. *Nat. Commun.* **8**, 407 (2017).

30. Nepal, R., Güngördü, U. &. Kovalev, A. A. Magnetic skyrmion bubble motion driven by surface acoustic waves. *Appl. Phys. Lett.* **112**, 112404 (2018).

31. Weiler, M. et al. Spin pumping with coherent elastic waves. *Phys. Rev. Lett.* **108**, 176601 (2012).

32. Davis, S., Baruth, A. & Adenwalla, S. Magnetization dynamics triggered by surface acoustic waves. *Appl. Phys. Lett.* **97**, 232507 (2010).

33. Camara, I. S., Duquesne, J.-Y., Lemaître, A., Gourdon, C. & Thevenard, L. Field-free magnetization switching by an acoustic wave. *Phys. Rev. Appl.* **11**, 014045 (2019).

34. Thevenard, L. et al. Precessional magnetization switching by a surface acoustic wave. *Phys. Rev. B* **93**, 134430 (2016).

35. Edrington, W. et al. SAW assisted domain wall motion in Co/Pt multilayers. *Appl. Phys. Lett.* **112**, 052402 (2018).

36. Yokouchi, T. et al. Creation of magnetic skyrmions by surface acoustic waves. *Nat. Nanotechnol.* **15**, 361−366 (2020).

37. Pollard, S. D. et al. Observation of stable Neél skyrmions in cobalt/palladium multilayers with Lorentz transmission electron microscopy. *Nat. Commun.* **8**, 14761. (2017).

38. Bottcher, M., Heinze, S., Egorov, S., Sinova, J. & Dupe, B. B–T phase diagram of Pd/Fe/Ir (111) computed with parallel tempering Monte Carlo. *New J. Phys.* **20**, 103014 (2018).

39. Chikazumi, S. *Physics of Ferromagnetism* 2nd edn, Ch.12 (Oxford Univ. Press, 1997).

40. Dreher, L. et al. Surface acoustic wave driven ferromagnetic resonance in nickel thin films: Theory and experiment. *Phys. Rev. B* **86**, 134415 (2012).

41. Sasaki, R., Nii, Y., Iguchi, Y. & Onose Y. Nonreciprocal propagation of surface acoustic wave in Ni/LiNbO$_3$. *Phys. Rev. B* **95,** 020407 (2017).

42. Shuai, J. T., Ali, M., Lopez-Diaz, L., Cunningham, J. E. & Moore, T. A. Local anisotropy control of Pt/Co/Ir thin film with perpendicular magnetic anisotropy by surface acoustic waves. *Appl. Phys. Lett.* **120**, 252402 (2022).

43. Küß, M. et al. Nonreciprocal Dzyaloshinskii–Moriya Magnetoacoustic Waves. *Phys. Rev. Lett.* **125,** 217203 (2020).

44. Weiler, M. et al. Elastically Driven Ferromagnetic Resonance in Nickel Thin Films. *Phys. Rev. Lett.* **106**, 117601 (2011).

**Methods**

To fabricate the delay line structure, we designed the IDTs with the width of 5 μm by using UV lithography followed by deposition of Ti (5 nm)/Al (100 nm) with electron beam evaporation. The distance of the two IDTs was 300 μm and each IDT had 70 pairs of single-type fingers. Here, the 128°-rotated, Y-cut LiNbO$_3$ substrate was selected for their piezoelectric property. Then, the magnetic films were patterned into the channel shape by standard optical lithography and lift-off process. The films were deposited at room temperature onto 5 mm × 5 mm LiNbO$_3$ substrate for magnetic property measurements via d.c. magnetron sputtering with a base vacuum better than $8.0 \times 10^{-5}$ mTorr, and the working argon pressure was 3 mTorr. The complete stack structure used in the measurements is LiNbO$_3$ substrate/Co(0.3 nm)/Pd(0.9 nm)/Co(0.3 nm)/Pd(0.9 nm)/Co(0.3 nm)/Pt(1.4 nm). The substrate temperature was monitored by an infrared camera after the RF power was applied for 2 minutes. The transmission spectrums between two IDTs were measured using a vector network analyzer and a RF signal generator was used to input the power during the experiments. The real-space skyrmions were observed using a polar MOKE microscope.

**Acknowledgements**

This work was supported by the National Key R&D Program of China (Grant No. 2022YFA1402603), the National Natural Science Foundation of China (Grant Nos. 52225106 and 51871130), and the Natural Science Foundation of Beijing Municipality (Grant No. JQ20010). C.S. acknowledges the support of Beijing Innovation Center for Future Chip (ICFC), Tsinghua University.